# Mapping how local perturbations influence systems-level brain dynamics

Leonardo L Gollo [a], James A Roberts [a,b], Luca Cocchi [a]

[a]QIMR Berghofer Medical Research Institute, Brisbane, Australia.
[b]Centre of Excellence for Integrative Brain Function, QIMR Berghofer Medical Research Institute, Brisbane, Australia.

Running title: Local perturbations and systems-level dynamics

Corresponding authors:

Leonardo L. Gollo, QIMR Berghofer Medical Research Institute, Brisbane, Australia. E-mail: leonardo.l.gollo@gmail.com

Luca Cocchi, QIMR Berghofer Medical Research Institute, Brisbane, Australia. E-mail: Luca.Cocchi@qimrberghofer.edu.au

Words: 9081 (without abstract and references but including figure legends)

Number of Figures: 9

SI: None

Conflict of interest: None

**Abstract**

The human brain exhibits a relatively stable spatiotemporal organization that supports brain function and can be manipulated via local brain stimulation. Such perturbations to local cortical dynamics are globally integrated by distinct neural systems. However, it remains unclear how and why local changes in neural activity affect large-scale system dynamics. Here, we briefly review empirical and computational studies addressing how localized perturbations affect brain activity. We then systematically analyze a model of large-scale brain dynamics, assessing how localized changes in brain activity at the different sites affect whole-brain dynamics. We find that local stimulation induces changes in brain activity that can be summarized by relatively smooth tuning curves, which relate a region's effectiveness as a stimulation site to its position within the cortical hierarchy. Our results also support the notion that brain hubs, operating in a slower regime, are more resilient to focal perturbations and critically contribute to maintain stability in global brain dynamics. In contrast, perturbations of peripheral regions, characterized by faster activity, have greater impact on functional connectivity. As a parallel with this region-level result, we also find that peripheral systems such as the visual and sensorimotor networks were more affected by local perturbations than high-level systems such as the cingulo-opercular network. Our results highlight the importance of a periphery-to-core hierarchy to determine the effect of local stimulation on the brain network. We also provide novel resources to orient empirical work aiming at manipulating functional connectivity using non-invasive brain stimulation.

**Introduction**

Brain functions depend on complex dynamic interactions between distinct brain regions that define neural systems (Sporns, 2011). This systems-level architecture and dynamics can be observed across different scales, from microscopic groups of cells (Oh et al., 2014; Schroeter et al., 2015; van den Heuvel et al., 2016; White et al., 1986) to macroscopically defined brain areas (Cocchi et al., 2013a; Hagmann et al., 2008; Sporns et al., 2005; Zalesky et al., 2014). Recent neuroimaging works have been instrumental in mapping the structural (Hagmann et al., 2008; van den Heuvel et al., 2012; van den Heuvel and Sporns, 2011) and functional (Gordon et al., 2014; Power et al., 2011; Yeo et al., 2011) architectures of the human brain at the macroscale. Functional magnetic resonance imaging (fMRI) data obtained while participants are not engaged in any specific task (resting-state) has been instrumental to understand the macroscopic network architecture supporting brain functions (Power et al., 2011; Yeo et al., 2011). While such functional architecture may be slightly renegotiated as a function of external and internal contexts (Bassett et al., 2011; Cocchi et al., 2013a; Cocchi et al., 2013b; Cole et al., 2013; Hearne et al., 2015; Leech et al., 2012), results from these studies have highlighted that the human brain is broadly organized into several large-scale neural systems that support diverse brain functions such as introspection, memory, task control, attention, and sensory processes (Power et al., 2011; Yeo et al., 2011).

An important unresolved question for neuroscience is how selective changes in neural activity in localized brain areas are globally integrated by the brain. Filling this knowledge gap is essential to understand the core principles supporting dynamic brain activity and functions. Recent neuroimaging studies have showed that changes in neural activity in a localized cortical region lead to selective and distinct changes in activity and functional connectivity within and between large-scale neural systems (Andoh et al., 2015; Cocchi et al., 2015; Eldaief et al., 2011; Grefkes and Fink, 2011; Johnen et al., 2015; Sale et al., 2015; Watanabe et al., 2013). The mapping of the whole-brain effects of local changes in neural activity is a recent field of research, and the degree of specificity of the observed global effects is unclear. Moreover, the neural principles that explain how a change in local neural activity may selectively alter its communication with other brain areas remain unknown. Such knowledge is necessary to understand how information circulates throughout the brain and how such flows can be altered. For example, emerging evidence suggests that neuropsychiatric pathologies such as obsessive-compulsive disorders (Cocchi et al., 2012b; Harrison et al., 2009), schizophrenia (Cocchi et al., 2014; Fornito et al., 2011; Zalesky et al., 2012), attention deficit hyperactivity disorder (Cocchi et al., 2012a) and melancholic depression (Hyett et al., 2015) are linked to abnormal interactions between brain areas. Such large-scale deregulations in functional connectivity may emerge from selective insults in highly interconnected core regions (Aerts et al., 2016; Crossley et al., 2014; Fornito et al., 2015; van den Heuvel et al., 2013), but the exact mechanisms leading to the putative propagation of local insults to large-scale deficits in system dynamics are not known.

Here, we start in Part I by briefly reviewing results from recent multimodal experiments assessing how a local change in neural activity can propagate and modulate large-scale brain dynamics. We deliberately focus our attention on research assessing how a non-invasive perturbation of local neural activity can alter the dynamics of the whole brain. Although they present obvious advantages such as exquisite spatial and temporal resolution, invasive studies assessing the widespread impact of altered local activity were omitted because of their inability to comprehensively assess whole-brain changes in neural system dynamics (Fox et al., 2014; Keller et al., 2011; Keller et al., 2014; Lega et al., 2015). Moreover, we do not focus on studies adopting techniques that have a relatively broad effect on regional brain activity, such as transcranial direct-current stimulation (tDCS) (Bikson et al., 2013), which is thought to modulate (rather than perturb) local neural activity [i.e., "sub-threshold" neuromodulation; (Zaghi et al., 2010)]. The remainder of Part I links this empirical literature with existing computational modeling work that investigates the putative mechanisms underpinning the effects of brain stimulation.

Building on the reviewed studies, Part II of this paper presents the results of an original investigation that uses computational modeling to assess putative principles explaining the effects of local perturbations on within- and across-systems neural dynamics. Results from this investigation yield a comprehensive map of how focal stimulation affects neural activity throughout large-scale brain systems. This mapping highlights the importance of brain organizational features such as the cortical hierarchy and the core-periphery axis to explain the impact of local changes in brain activity to whole-brain dynamics.

**Part I: An overview on the system-level effects of local changes in brain activity**

*Empirical studies mapping changes in whole-brain connectivity following non-invasive perturbations to local neural activity*

The interest in transcranial magnetic stimulation (TMS) was originally driven by two main applications: i) using brain stimulation to study the specialized functions of spatially segregated cortical regions; and ii) the potential use of such techniques to restore brain functions after local neural insults such as stroke. Thanks to recent advances in the fields of brain stimulation and neuroimaging, such techniques have started to be combined to assess how segregated brain areas are functionally coupled and how such coupling can be altered by focal changes in neural activity [(Sale et al., 2015), but see some early pioneering PET works; e.g., (Fox et al., 1997)].

An emerging body of work combines fMRI and/or electroencephalography (EEG) with non-invasive brain stimulation (e.g. TMS) to assess large-scale neural dynamics, as reviewed recently elsewhere (Bortoletto et al., 2015; Sale et al., 2015). TMS offers a range of stimulation paradigms, with varied local neural effects. Common stimulation protocols such as theta burst stimulation (TBS) can excite or inhibit neural activity in local cortical areas by changing the frequency of magnetic bursts during stimulation (Huang et al., 2005). Behaviorally, the excitatory or inhibitory nature of TMS is often identified from observed motor outputs: excitation is identified as a reduction in the intensity of stimulation needed to induce a movement whereas inhibition results in an increased motor threshold.

Studies combining fMRI and TMS have shown that focal TMS in a cortical area can modulate the focal activity in other brain regions and the related behavior (Bestmann et al., 2005; Ruff et al., 2006). In their seminal study, Ruff et al. (2006) showed that stimulation of the frontal eye fields (FEF) during visual tasks performance enhances neural activity in distal sensory visual areas encoding peripheral stimuli. This change in remote brain activity following TMS was paralleled by an enhancement in the detection of visual stimuli presented in the periphery relative to the fovea (Ruff et al., 2006). These remote, yet circumscribed, changes in brain activity suggest a change in the patterns of communication between the cortical area targeted by TMS (FEF) and early visual regions. In support of this hypothesis, recent investigations have shown that focal changes in neural activity via TMS selectively alter patterns of functional and effective connectivity in related brain systems such as the sensorimotor (Grefkes and Fink, 2011; Grefkes et al., 2010) and the default mode (Eldaief et al., 2011) networks. Further work indicates that although local TMS affects patterns of connectivity within the system containing the target region, local TMS may also change between-system interactions (Cocchi et al., 2015; Valchev et al., 2015; Watanabe et al., 2014). For example, we recently showed that selective inhibition of the right primary motor cortex enhanced integration between areas composing the sensorimotor system and reduced the communication between that system and other systems of the brain Cocchi et al. (2015). Interestingly, such changes in dynamics occurred in the context of preserved patterns of functional connectivity between highly interconnected brain areas forming the *rich club* (Collin et al., 2013; van den Heuvel et al., 2012; van den Heuvel and Sporns, 2011). These results suggest that a local change in neural activity may be gradually integrated by changes in selective network dynamics, preserving the core connectivity patterns supporting the bulk of information transfer in the brain.

Yet, the neural principles supporting the observed effects of focal TMS on brain systems remain elusive. Work combining structural and functional neuroimaging techniques with brain stimulation suggests that the selective effect of local TMS on brain systems could be associated, at least in part, to patterns of structural connectivity between brain regions (Andoh et al., 2015; Grayson et al., 2016). However, it remains unclear to what extent the underlying anatomical connections determine changes in functional brain networks. Until it is known which factors predict the global impact of a focal change in brain activity, the planning and interpretation of hypothesis-driven TMS experiments will remain challenging. In this context, computational modeling can provide a more principled framework for mapping of the whole-brain effects of local TMS, and provide important clues toward the putative neural mechanisms explaining the system-level effects of targeted changes in local brain activity.

*Toward a mechanistic understanding of global changes in system dynamics following focal changes in neural activity*

Computational models are crucial for linking the various independent experimental results into a unified theoretical framework, and for making predictions that can be tested in future experiments (Breakspear et al., 2010b). For the macroscopic scale of the whole brain the most suitable models are mean-field (or *neural mass* or *neural field*) models. Such models describe the aggregate behavior of a very large numbers of neurons (Deco et al., 2008). This is a natural level of description for neuroimaging data. In fact, for whole brain recordings, existing noninvasive technologies restrict us to measuring the bulk activity of neural masses comprising a huge number of neurons (Foster et al., 2016). This constraint is true also for most stimulation techniques, where a macroscopic electrode (or other stimulation device) affects the activity of a large number of neurons simultaneously (Allen et al., 2007). Thus, we restrict our attention here to models suitable for whole-brain descriptions of neural dynamics and stimulation protocols. For reviews of macroscopic neural modeling more broadly, we direct the reader to existing reviews (Coombes, 2010; Deco et al., 2008).

Mean-field models at the whole-brain scale roughly fall into two categories: coupled neural masses (Gollo and Breakspear, 2014; Sanz-Leon et al., 2015; Schmidt et al., 2015) and neural fields (Atasoy et al., 2016; Jirsa and Haken, 1996; Robinson et al., 2001; Robinson et al., 2016). Neural mass models describe the dynamics of a spatially-localized large number of neurons. To describe whole-brain dynamics these neural masses are coupled together, thus yielding a discrete approximation to the brain at a macroscopic resolution (typically ~30-1000 nodes). Neural field models, on the other hand, use a continuum formulation where the spatially-extended cortex is treated as a continuous object. While analyses on idealized geometries are tractable analytically, other approaches (particularly on realistic cortical geometries) require numerical simulations, for which the continuum must be discretized. In this sense the two approaches are largely equivalent in the limit of fine spatial resolution, depending on the nature of the coupling. Recent approaches have been proposed attempting to bridge the discrete and continuous descriptions (Robinson, 2005; Spiegler et al., 2016).

The macroscopic local effects of non-invasive brain stimulation (specifically TMS) have been modeled recently in two related mean-field approaches. One approach treats TMS as modifying local activity via spike-timing dependent plasticity (Fung et al., 2013; Wilson et al., 2014). Although this specific mechanism is phenomenological, the models enable prediction of long-term responses to common TMS protocols. The second approach includes a more physiologically-realistic description of calcium-dependent plasticity, thus showing how the system-level plasticity driven by local TMS emerges from microscopic mechanisms (Fung and Robinson, 2013, 2014; Wilson et al., 2016a; Wilson et al., 2016b). These neural field models naturally describe spatially-extended cortex, but to date only the spatially-uniform effects of TMS perturbations have been explored in this framework. Similarly to the studies of lesions (Aerts et al., 2016; Alstott et al., 2009; Arsiwalla et al., 2015; Hagmann et al., 2008), analysis of how local perturbations affect dynamics across the whole brain will likely be fruitful.

Biophysical models of local TMS effects across brain regions have thus far been restricted to only small numbers of regions. The acute effects of TMS (within the first ~200 ms) have been modeled using three coupled neural masses to represent three distant interconnected brain regions (Cona et al., 2011). These authors also fitted their model to simultaneous EEG data to infer the directionality of inter-region connectivity. However, this approach is difficult to extend to the whole-brain scale. A coupled-neuron model has also been used to study propagation of TMS pulses between three cortical areas, using >30,000 neurons in each area (Esser et al., 2005). This is of course not a mean-field model, and its suitability to the study of whole-brain dynamics is limited due to the exceedingly high computational requirements when simulating so many neurons.

Spatial effects of stimulation have also been modeled in other stimulation modalities. EEG event-related potentials (ERPs) evoked by sensory stimuli have been modeled in neural field theory, including their spatial propagation away from the localized stimulus (Rennie et al., 2002). In this model, which forms part of the broader neural-field theory whose extension to TMS was discussed above (Fung and Robinson, 2013, 2014), the ERP is analytically tractable and attentional effects can be described by plausible shifts in parameters. Spatial ERP effects have also been explored using coupled neural masses in the context of dynamic causal modeling (David et al., 2006), with the aim of inferring effective connectivity between connected brain regions. These last two studies are examples of using coupled biophysical mean-field models to study large-scale brain activity.

Another approach is to use more abstract models to infer the spread of activity within the whole-brain structural network. For example, spreading dynamics have been explored using an idealized model akin to those used to model the spread of infections on social networks (Mišić et al., 2015). This revealed the role of hubs and backbones in spreading activity across the brain, and the role of associative areas in integrating competing cascades of activity. The global "dynamics" in this case are essentially directly endowed by the structural connectivity without incorporating any local dynamics. To eliminate this limitation without adding all the complexity of a biophysical model, another approach is to employ simplified local dynamics using idealized mathematical models. An influential choice that incorporates cortical oscillatory dynamics is the Kuramoto model (Rodrigues et al., 2016), which we will discuss in detail in Part II. Another is the use of phenomenological "normal form" models, which capture the essence of some aspect of nonlinear dynamics (e.g., bifurcations to oscillations), but with relatively few parameters. Such models have recently been used in conjunction with structural connectome data to systematically explore the spatial patterns that emerge following focal stimulation (Spiegler et al., 2016). Together, these studies provide additional insights into the putative neural mechanisms that support the propagation of information within the brain. Specifically, they highlight the role of structural connectivity in constraining the spread of the dynamics (Honey et al., 2007; van den Heuvel et al., 2013).

An extreme example of the spread of activity evoked by a localized stimulus to encompass the whole brain is seizures triggered by a stimulus. This is of relevance to TMS studies: the range of safe stimulation parameters is constrained by the requirement to not induce seizures through over-excitation (Rossi et al., 2009). The focal initiation and subsequent spread of seizures have been studied in spatially-extended neural models (Goodfellow et al., 2012; Kim et al., 2009). Spreading of seizures from a focal onset is now widely thought to be fundamentally a network phenomenon, such that patient-specific modeling of seizure propagation may open new therapeutic possibilities (Proix et al., 2016). Mapping the safe range of stimulation parameters is a potentially fruitful avenue for modeling studies to guide future experiments, and indeed stability has been studied in neural field models (Fung and Robinson, 2013; Wilson et al., 2016b). More broadly, the recent interest in controllability of brain dynamics (Betzel et al., 2016) will likely drive increased interest in noninvasive stimulation methods.

**Part II: Modeling the global effects of focal changes in neural activity**

In this second part we present the results of a systematic exploration of the effects that focal perturbations in neural activity across the human cortex may have on widespread brain systems (i.e., resting-state neural networks). The central aim of our modeling work is to determine the structural and functional principles that underpin how localized perturbations to brain activity modulate large-scale dynamics. Specifically, our simulations: (i) seek a general pattern of whole-brain (global) changes in functional connectivity as a function of the connectivity properties of the targeted region; (ii) assess the role of a cortical structural and functional hierarchy in determining the changes in whole-brain functional connectivity; and (iii) investigate how functional connectivity within systems changes in response to local perturbations in brain activity.

*Model*

We used a minimal and standard computational model of synchronization, the Kuramoto model (Kuramoto, 1984). The relative simplicity of this model allows the systematic exploration of focal perturbations across the whole cerebral cortex. The model captures essential aspects of macroscopic dynamics that appear in more complex models (Bhowmik and Shanahan, 2013) and has been shown to reproduce changes in functional connectivity following local TMS in real fMRI data (Cocchi et al., 2016). TMS changes the amplitude (Allen et al., 2007) and frequency (Okamura et al., 2001; Thut and Pascual-Leone, 2010) of local oscillations, which is consistent with the results of our recent work (Cocchi et al., 2016). Thus, the Kuramoto model provides a balanced trade-off between complexity and plausibility for whole-brain functional connectivity, incorporating key anatomical constraints and modeling major dynamical elements captured with neuroimaging techniques such as fMRI (Hellyer et al., 2015; Lee et al., 2016).

We modeled slow (0.01-0.1 Hz) fluctuations of the blood-oxygen-level dependent (BOLD) signal in 513 volumetrically similar brain regions as oscillators. Both fast (Breakspear et al., 2010a) and slow (Schmidt et al., 2015) cortical oscillations have been successfully modeled with systems of coupled phase oscillators. The dynamics of the phase $\theta_i(t)$ corresponding to the oscillator of each region $i$ is given by (Kuramoto, 1984),

$$\frac{d\theta_i}{dt} = \omega_0^i + \lambda \sum_{j=1}^{N=513} W_{ij} \sin(\theta_j - \theta_i), \qquad (1)$$

where $\omega_0^i$ is the intrinsic frequency of region $i$, $\lambda= 0.0028$ is a constant previously determined by comparison with resting-state fMRI data (Cocchi et al., 2016), and $W_{ij}$ is the normalized weight of the anatomical connection between regions $i$ and $j$ extracted from a group-averaged whole-brain structural connectivity matrix $W$ (Roberts et al., 2016). A brief overview of this modeling approach is presented in Fig. 1. Adopting an empirically obtained structural connectivity and a distribution of natural frequencies, the dynamics of the whole brain is obtained integrating Eq. (1) to yield time series for each cortical region. Simulations are run for a duration of 10 min and repeated 50 times with different random initial conditions. Discarding the first two minutes, we compute the mean functional connectivity matrix across trials. This connectivity matrix is then partitioned into functional systems as previously proposed in the literature by Power et al. (2011).

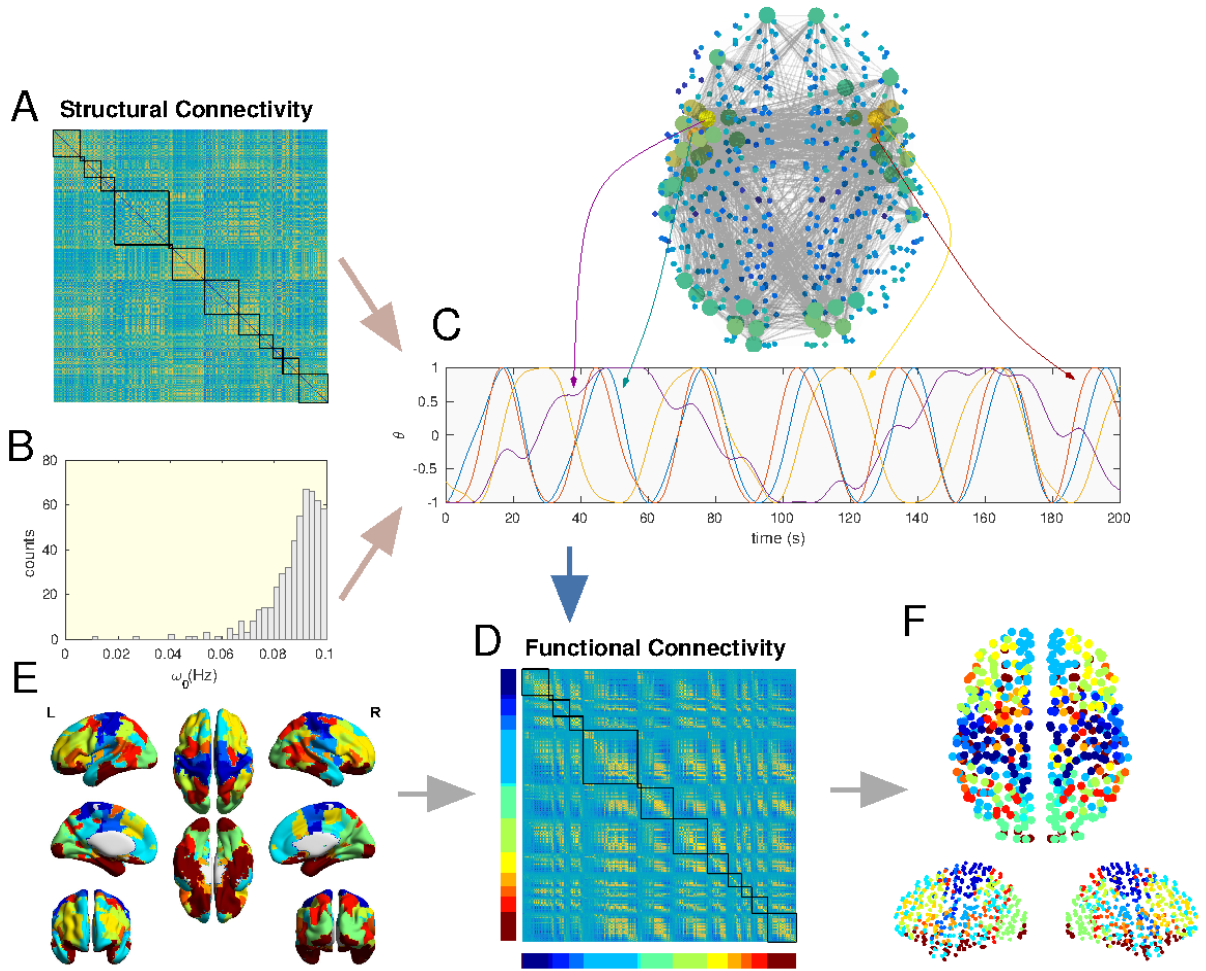

**Figure 1.** *Whole-brain model description.* A. Structural connectivity matrix determined using probabilistic tractography in 75 healthy participants (details in Roberts et al. (2016)). The connectome was characterized using a 513-node brain parcellation. Warmer colors indicate stronger connections. B. Distribution of natural frequencies, which are determined by the connectivity strength (sum of weights from the structural connectivity) of each region (Eq. 2), and tuned to best replicate resting-state patterns of connectivity observed in real data (Cocchi et al., 2016). C. Rich-club connectivity (top; large circles correspond to hubs, gray lines the hub-to-hub connections), and illustrative time series of a few rich club and peripheral regions (bottom) obtained from model. D. Functional connectivity matrix obtained in the model. Warmer colors indicate stronger functional connectivity. Color bar represents the different functional systems. E. Functional brain systems isolated using real resting-state fMRI data (Power et al., 2011) and colored according to the color bar in panel D. F. Centroids of the 513 brain regions used in the model color-coded by the functional systems of panel E.

*Anatomical connectivity*

As described in our previous work (Cocchi et al., 2016), the representative connectivity matrix *W* used in the Kuramoto model (Equation 1) was obtained from a sample of 75 healthy adult participants (Roberts et al., 2016). Densely-seeded probabilistic tractography was used to estimate the group-average number of white matter tracks connecting all pairs of brain regions [513 x 512= 262,656 connections; details in Roberts et al. (2016)]. The resulting structural connectivity matrix (connectome) was normalized to have a maximum weight of 1. Figure 1 depicts the connectivity matrix as well as the rich club representation of the connectivity between the hubs (i.e., top 75 cortical regions that have strength of at least one standard deviation above the mean).

*A hierarchy of intrinsic timescales across the cortex*

Based on previous knowledge of the hierarchy of timescales across the primate cortex (Chen et al., 2015; Gollo et al., 2015; Hasson et al., 2015; Honey et al., 2012; Murray et al., 2014), the intrinsic values of $\omega_0$ are assumed to be a function of the anatomical node strength $s_i = \sum_{j=1}^{513} W_{ij}$. Specifically, the intrinsic frequency for each region is given by:

$$\omega_0^i = a - (a - b)\left(\frac{s_i - s_a}{s_b - s_a}\right)^2, \qquad (2)$$

where $a = 0.1$ Hz and $b = 0.01$ Hz are the maximum and minimum oscillatory frequencies, and $s_a = \min(s)$ and $s_b = \max(s)$ are the corresponding maximum and minimum strengths, respectively. A histogram of the natural frequencies is shown in Fig. 1B. The distribution has a mean value of $\langle \omega_0 \rangle = 0.0883$ Hz. Hence, the cortical hierarchy (Felleman and Van Essen, 1991; Maunsell and van Essen, 1983; Van Essen and Maunsell, 1983) maps into a gradient of timescales (Murray et al., 2014). The distribution of $\omega_0$ was tuned to closely reproduce the resting state functional connectivity (0.01-0.1 Hz) obtained using fMRI (Cocchi et al., 2016). The parameters *a* and *b* were then chosen to capture the outer extremes of the BOLD signal frequency bandwidth. Figure 2A illustrates the natural frequencies $\omega_0$ as a function of the cortical region indices ordered by their anatomical connectivity strength. After coupling the regions together, their observed oscillatory frequencies change due to network interactions: slower hub regions speed up, and faster peripheral regions slow down their activity (Fig. 2B&C). Coupling thus shifts the mean oscillatory frequency (across trials) of each region (Figure 2C). In addition to constraining the distribution of natural frequencies (Eq. 2), we have also calibrated the model (varying $\lambda$) to closely match the baseline functional connectivity (Cocchi et al., 2016).

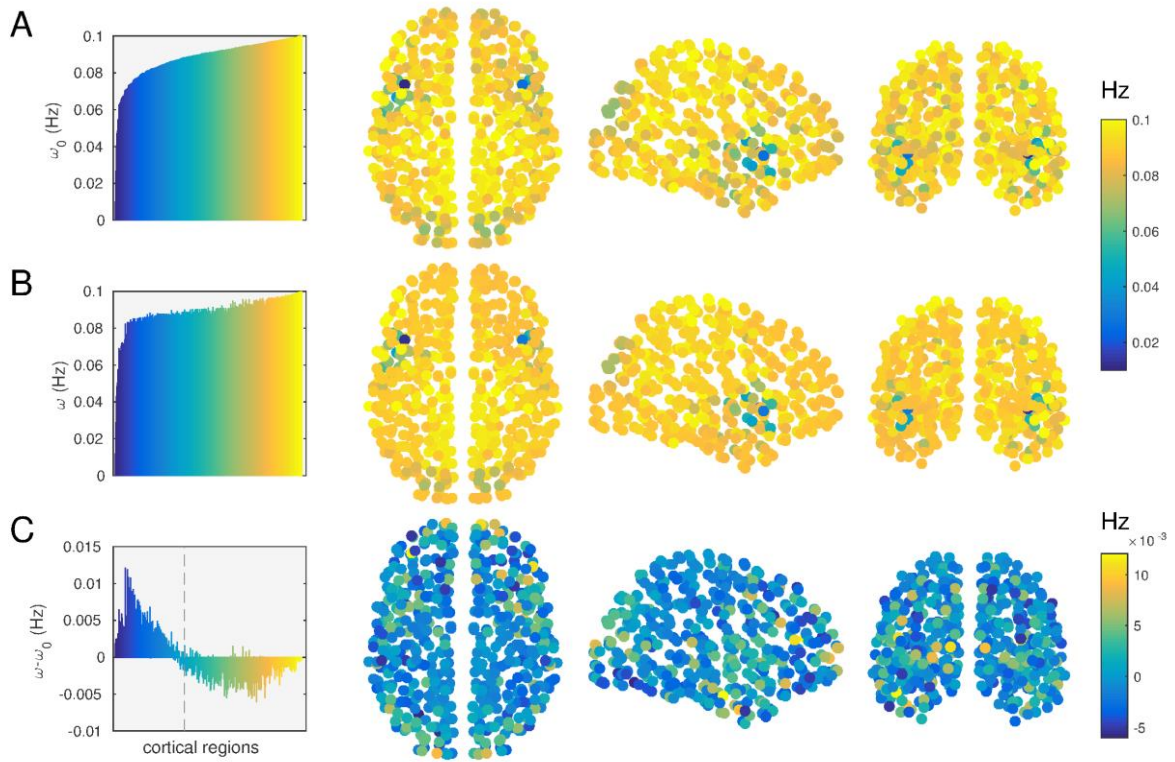

**Figure 2.** *Regional differences in resting-state oscillatory frequencies*. A. Natural frequencies of cortical regions plotted (left) against cortical region indices (ordered by their strength of anatomical connectivity), and shown in physical coordinates. Colors of brain regions denote $\omega_0$. B. Oscillatory frequencies after coupling regions together (mean across trials). Colors of brain regions denote $\omega$. C. Mean change in oscillatory frequency due to coupling. Dashed gray line denotes the cortical region with natural frequency $\omega_0^i = \langle \omega_0 \rangle$. Colors pf brain regions denote the coupling-induced frequency shift $\omega - \omega_0$.

*Brain stimulation*

Transcranial magnetic stimulation (TMS) alters the frequency and power of signals detected with neuroimaging techniques (Okamura et al., 2001; Thut and Pascual-Leone, 2010). Such effects are incorporated in our model by a change in the natural frequency of the stimulated region compared to the baseline (Eq. 2). Specifically, inhibitory stimulation is modeled as a reduction in the natural frequency $\omega_0^i$ of the stimulated region, and excitation is modeled as an increase in $\omega_0^i$ (Cocchi et al., 2016).

**Results**

We simulate whole-brain dynamics using the Kuramoto model. The model introduces heterogeneous natural frequencies across brain regions; in the dynamics that result, the observed frequencies are shifted slightly due to network interactions (Fig. 2C). This heterogeneity is governed by Eq. 2, which incorporates a simple relation determining the natural frequencies as a function of the strength of the anatomical connections that every region has with other brain regions. This implies a direct link between the anatomical hierarchy and the intrinsic oscillatory dynamics (Kiebel et al., 2008). In this framework, peripheral regions exhibit fast dynamics whereas brain hubs work in slower regimes (Gollo et al., 2015). To characterize the changes in brain dynamics, we then systematically stimulate each brain region (N = 513) in a series of numerical experiments. Such a thorough exploration of the widespread effects of the target-dependent local changes in neural activity is experimentally challenging. Our findings may therefore provide valuable insights to guide future hypothesis-driven experimental work.

*Tuning curves: Effects of local inhibition or excitation on whole-brain dynamics*

We find that changes in functional connectivity between the stimulated area and the rest of the brain (ΔFC) can vary substantially depending on the targeted brain region. Despite these heterogeneous effects, the changes in functional connectivity following local stimulations follow a characteristic pattern that can be viewed as a tuning curve (Fig. 3). By ordering the regions according to their strength of anatomical connectivity and intrinsic frequency of oscillation, we find two types of responses: regions show either an increase or decrease in functional connectivity following perturbations. The propensity of nodes to increase or decrease connectivity is determined by the natural frequency of each region with respect to the mean natural frequency. Strongly connected regions have slow timescales and weakly connected regions have fast timescales. Such regions intersect at the strength that corresponds to the mean natural frequency (vertical dashed line in Fig. 3). Moreover, local inhibition and excitation have similar but opposing tuning curves (with an anticorrelation of r = -0.6, Fig. 3A&B). Local inhibition increases functional connectivity in regions with low anatomical connectivity. In more interconnected core brain regions, inhibition results in decreases in functional connectivity. The opposite results are observed for local excitation of a cortical region. In both cases the regions with the very strongest and weakest connectivities are largely insensitive to the simulation.

Mapping the spatial distribution of the observed changes in functional connectivity following local inhibition reveals a rather complex pattern with small islands of strong increases or decreases in correlations and an absence of a clear structure (Fig. 3 C&D). We also find that, varying the strength of the stimulation ($\Delta\omega_0$) does not affect the shape of the tuning curves, but rather determines the amplitude of the changes in functional connectivity (Fig. 3E). These curves are smooth and show only the main trend of the brain regions with a similar strength in structural connectivity. In addition, the changes in functional connectivity corresponding to the node that shows the largest response to stimulation grow nonlinearly with the amplitude of stimulation: faster for weak stimulations and slower for strong stimulations (Fig. 3F). That is, at the regional level, changes in the intensity of stimulation can have a large impact. Indeed, the maximum changes in functional connectivity (Fig. 3F) refer to different regions for the different intensity values of $\Delta\omega_0$. At the single-region level, when the intensity of the stimulation exceeds a certain threshold changes in functional connectivity between the stimulated region and the rest of the brain may reverse [e.g. from an increase in positive correlations to a decrease in positive correlations, Cocchi et al. (2016)]. This intensity-dependent effect at the regional level highlights the importance of considering region-specific anatomical and functional profiles to predict the impact of a given stimulation to the brain network.

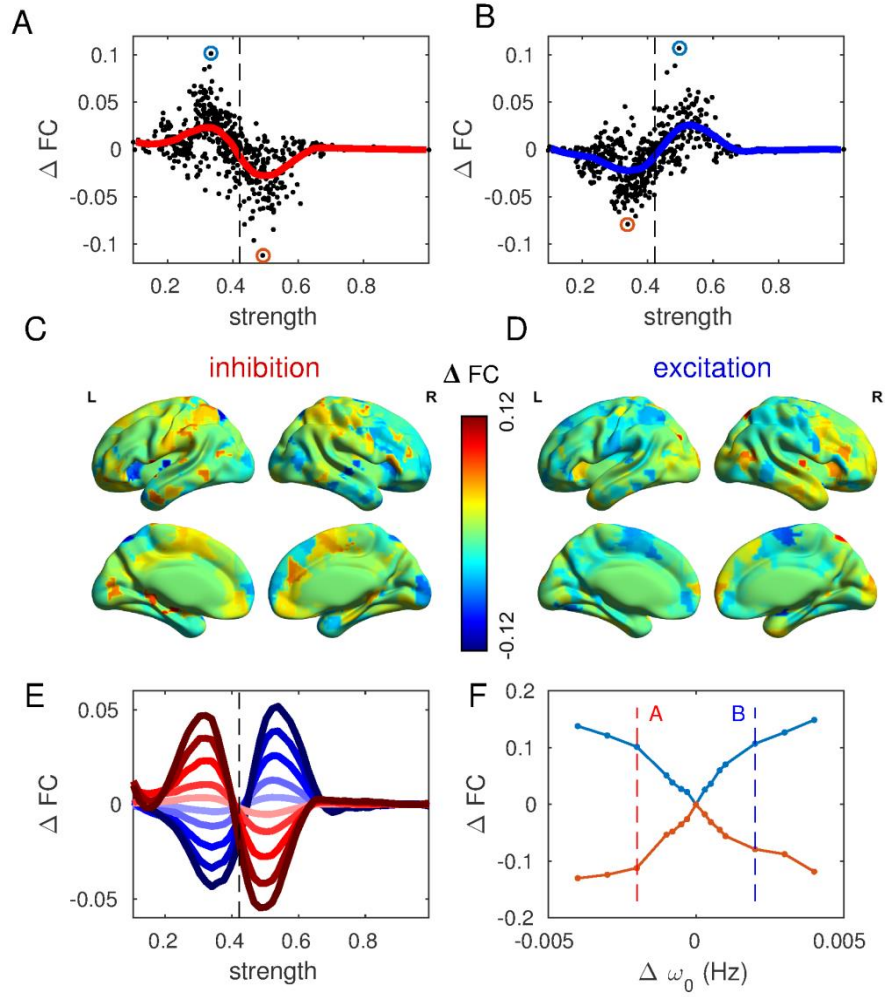

**Figure 3.** *Tuning curves summarizing the effects of local perturbations on whole-brain functional connectivity*. A. Tuning curve showing the change in functional connectivity $\Delta FC$ (black dots) versus the stimulated node's connectivity strength following selective inhibitory stimulation in each region of interest, with stimulation strength $\Delta \omega_0$ = -0.002 Hz. B. Tuning curve following excitatory stimulation, $\Delta \omega_0$ = 0.002 Hz. The vertical black dashed line indicates the node strength that corresponds to the mean natural frequency. Solid lines indicate the nonoverlapping sliding window mean, smoothed with the Savitzky-Golay filter (with length 15 and order 3). C and D. Spatial distribution of the strength of changes in functional connectivity following local stimulations. The color axis indexes the mean changes in functional connectivity $\Delta FC$ between the stimulated region and the rest of the brain. E. Tuning curves for different values of stimulation strength, $\Delta \omega_0$= -0.004, -0.003, -0.002, -0.001, -0.0003, 0.0003, 0.001, 0.002, 0.003, 0.004 Hz. Dark red represents strong inhibition and dark blue represents strong excitation. F. Maximal regional increase (blue) and decrease (orange) in the mean functional connectivity for different level of inhibition and excitation. Vertical dashed lines denote the corresponding extremal values of $\Delta FC$ in panels A and B (circled there in blue and orange).

One of the consequences of the proposed hierarchy of timescales is that regions with similar white matter connectivity strength have similar timescales. Hence, regions with similar strength are expected to show similar dynamical features. This is of course a major simplification because all of the connectome's fine-level complexity is overlooked and a single value (strength) describes each region. Nonetheless, our results show that such a reduction remains able to capture some fundamental features of whole-brain functional connectivity. Moreover, assessing the impact of focal changes in nodal activity shows a clear response trend from peripheral to core regions (Fig. 4). This specific result highlights the importance of the periphery-core axis to predict the effect of local stimulation to the whole brain functional connectivity. Our findings are also in line with the hypothesis of a rostro-caudal axis within the frontal cortex [green dashed line, Fig. 4 (right), (Badre and D'Esposito, 2009; Fuster, 2001)]. However, we find that a general rostro-caudal axis does not explain the behavior in other regions, exhibiting a relatively flat dependence on spatial position along the axis for regions outside the most rostral. We thus contend that the core-periphery topological axis provides a more general explanation for the ability of regions to affect large-scale dynamics, because it spans the entire brain.

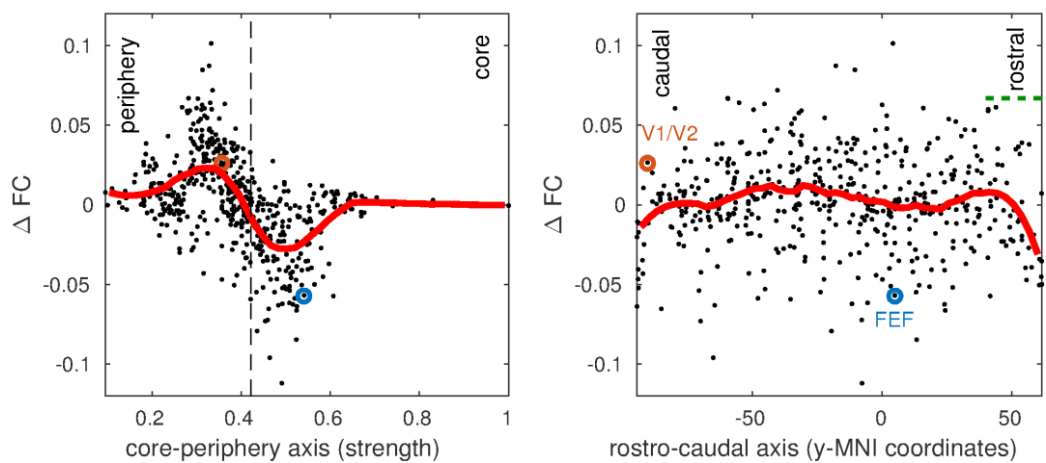

**Figure 4**: *Comparison between periphery-core and rostro-caudal organizational principles after inhibitory stimulation*. The effects of local stimulation are more evident when brain regions are organized along a periphery-core axis. A rostro-caudal organization reveals a specific trend only for the frontal brain regions (dashed green bar). Regions V1/V2 and FEF that have been confirmed experimentally are highlighted (Cocchi et al., 2016).

*Tuning curves: Magnitude of the impact of local changes in brain activity to whole brain connectivity*

Analysis assessing the impact of local stimulation on the whole brain network suggests that a relatively simple nonlinear function accounts for the relationship between anatomical and functional connectivity. However, this result explains only the average change in functional connectivity between the stimulated region and the rest of the brain. Instances of $\Delta FC \approx 0$ could mask cases in which balanced large increases and decreases in functional connectivity coexist but essentially cancel each other out. Indeed we found that several regions show a very small value of $\Delta FC$ for seed-to-whole-brain effects because their changes in positive and negative correlations are largely balanced, yielding a relatively small net effect. Figure 5 illustrates the case of one exemplar region that shows both large increases as well as decreases in functional connectivity with the rest of the brain. These effects are balanced, and hence $\Delta FC \approx 0$ for inhibition and excitation (Fig. 5B).

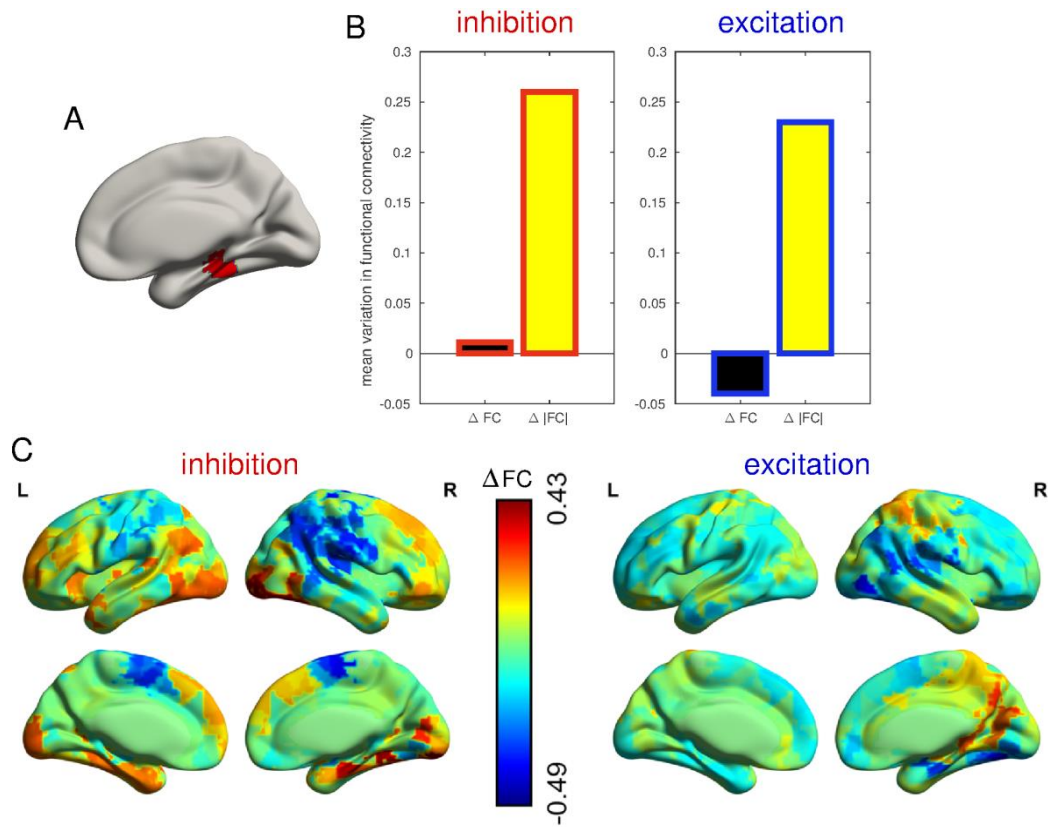

**Figure 5.** *Exemplar region showing large and balanced changes in functional connectivity following stimulation.* A. Location of the region targeted with stimulation (red). B. Mean and absolute changes

in functional connectivity following local stimulation ($\Delta\omega_0 = -0.002, 0.002$ Hz) of the region presented in panel A. C. Changes in functional connectivity between the stimulated area and the rest of the brain (i.e., with each of the other 512 regions).

Since the seed-to-whole-brain effects can be masked by similar increases and decreases in functional connectivity, we also analyze the mean of the absolute value of the changes in functional connectivity: $\Delta|FC|$. In analogy with $\Delta FC$ (Fig. 3), $\Delta|FC|$ can also give rise to tuning curves (Fig. 6). Instead of mirroring one another, tuning curves for $\Delta|FC|$ are very similar for both inhibition and excitation. As shown in Fig. 6A, peripheral regions show greater changes in their connectivity profile following focal stimulation compared to core regions. In contrast to $\Delta FC$ (Fig. 3), brain regions with natural frequency near the mean frequency $\langle\omega\rangle$ (i.e., with strength given by the vertical black dashed line, Fig. 6) show a relatively large value of mean $\Delta|FC|$. This result implies that perturbations in such regions have an impact on widespread connectivity that is relatively large in magnitude but opposite in sign.

For weak stimulation, the curves generated using absolute changes in functional connectivity are very similar for both the excitatory and inhibitory stimulation (Fig. 6A, right). However, as the intensity of the stimulation increases the excitatory and inhibitory tuning curves dissociate for peripheral and core regions. This dissociation occurs because strong local stimulation saturates the absolute changes in functional connectivity. Upon excitation, fast peripheral regions become even faster and their effect on functional connectivity decreases with stimulation intensity. On the contrary, strong inhibition of the same peripheral regions increases the functional connectivity of these regions with other brain areas operating at a significantly slower regime. A dissociation of excitatory and inhibitory effects occurs for core regions as the intensity of stimulation grows by a substantially smaller amount because the effect of stimulation is reduced in these hubs.

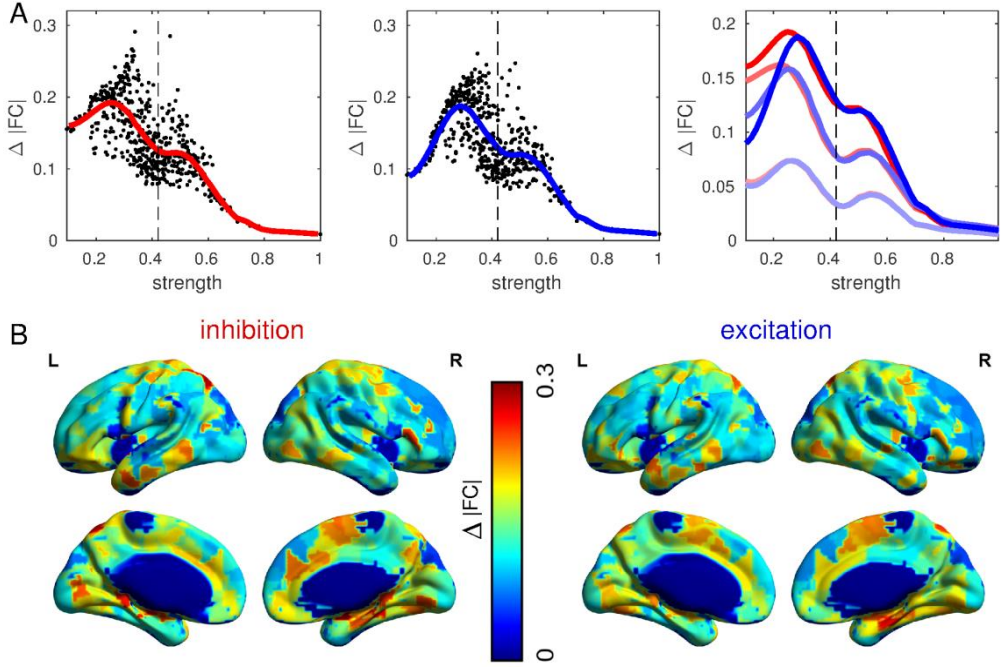

**Figure 6**. *Magnitude of the impact of local changes in brain activity to whole brain connectivity*. A. Tuning curves plotting the absolute value of the changes in functional connectivity $\Delta|FC|$ (black dots) between the stimulated region and the rest of the brain against the connectivity strength of the stimulated region. Curves are smoothed as in Fig. 3. Left panel, local inhibition with $\Delta\omega = -0.002$ Hz; middle panel, excitation with $\Delta\omega = 0.002$ Hz. The right panel shows the tuning curves for varying $\Delta\omega = -0.002, -0.001, -0.0003$ (inhibition); $0.0003, 0.001, 0.002$ (excitation) Hz. Dark red represents strong inhibition and dark blue represents strong excitation. B. Spatial distribution of brain regions showing magnitude of changes in functional connectivity as a function of the strength of anatomical connectivity.

Figure 6B shows how the mean absolute values of the changes in functional connectivity are distributed in the brain. The mean change of $\Delta|FC|$ across all regions is very similar, $0.142 \pm 0.052$ for inhibition and $0.140 \pm 0.049$ for excitation. Moreover, the magnitude of diffuse changes following each region's excitation or inhibition are highly correlated (r = 0.8). This result suggests that some highly reactive regions cause larger effects in functional connectivity following both inhibition and excitation. Hence, these regions should be the first target regions for the next validation experiments.

The tuning curves show either a peak and a trough (Fig. 3) or two peaks (Fig. 6). Both types of tuning curves have the mean natural frequency poised at a point of weak response between the peaks/troughs of stronger responses. This means that certain brain regions with natural frequency near (but not too close) to the average of the whole-brain ($\langle\omega_0\rangle$) are more likely to increase/decrease their functional connectivity with other regions. Regions that are far from the mean natural frequency (i.e., close to the extremes of too fast or too slow) only drive small effects. And regions that operate close to the brain's mean-frequency regime can show a balanced effect (with equal-but-opposite increases and decreases) in functional connectivity. By projecting the values of functional influence onto the cortical surface, this study provides a systematic mapping of the effects of inhibition and excitation on large-scale resting-state brain dynamics.

*Effects of local changes in brain activity on specialized functional brain systems*

Here we answer the following: how do manipulations of local brain activity shape the patterns of functional connectivity within and between specialized functional systems? To address this question we first quantify the extent to which functional connectivity within functional systems can be changed following stimulation of each individual brain region. To this end, the brain network is initially divided into twelve known functional resting-state brain systems [Fig. 1; Power et al. (2011)]. These systems exhibit specific anatomical and functional properties. For example, each system comprises a defined number of regions and a specific pattern of anatomical connectivity (i.e., connectivity strength) within itself and with other systems.

We then stimulate each region and measure the mean absolute value of the changes in functional connectivity between this region and all other regions located within the functional system of interest. To quantitatively explore the effects of local changes in brain activity within and outside the system to which the stimulated region belongs, we divide the regions composing the whole brain into two groups: (i) internal regions defining each system (green in Fig. 7), and (ii) external regions that belong to other functional systems (purple in Fig. 7). We term the mean effect after stimulating a whole system the *functional influence*. Results showed that the functional influence is consistent across systems for the external regions (Fig. 7A, purple bars) but vary substantially for the internal regions (green bars). Specifically, we found that peripheral systems (having a lower mean strength in the structural connectivity) show larger functional influence. Conversely, systems defined by larger mean strength in anatomical connectivity had a lower functional influence (Fig. 7B). In fact, the within-system functional influence is anticorrelated with the mean strength of functional systems (r = -0.45) and with their strongest constituent region (r = -0.55). Moreover, our finding shows that larger functional influence occurs for peripheral systems, mirroring the stronger stimulation-induced effects on functional connectivity found for peripheral regions (Fig. 6). This indicates a general greater functional flexibility of both peripheral regions and systems.

Figure 7C shows the functional influence of specific brain regions. Purple areas are located outside the system of interest (depicted in green) whereas green areas are located within the system of interest. Darker purple and green colors represent areas that induce larger changes in functional connectivity within the system of interest. We find that changes in within-system dynamics are contingent on the accuracy of the localization of the site of stimulation, both within and outside the system of interest. Stimulation of some cortical regions is able to alter functional connectivity within the targeted system whereas the stimulation of other cortical regions may have little or no effect. The spatial distribution of the effects does, however, suggest a degree of continuity in the functional influence of each region so that neighboring regions generally show similar effects. All networks show internal and external *hotspots* able to induce substantial changes in functional connectivity. Nonetheless, internal hotspots (dark green in Fig. 7B) are dominant in peripheral systems such as vision and somatosensory networks compared to high-order associative systems (e.g., salience and cingulo-opercular).

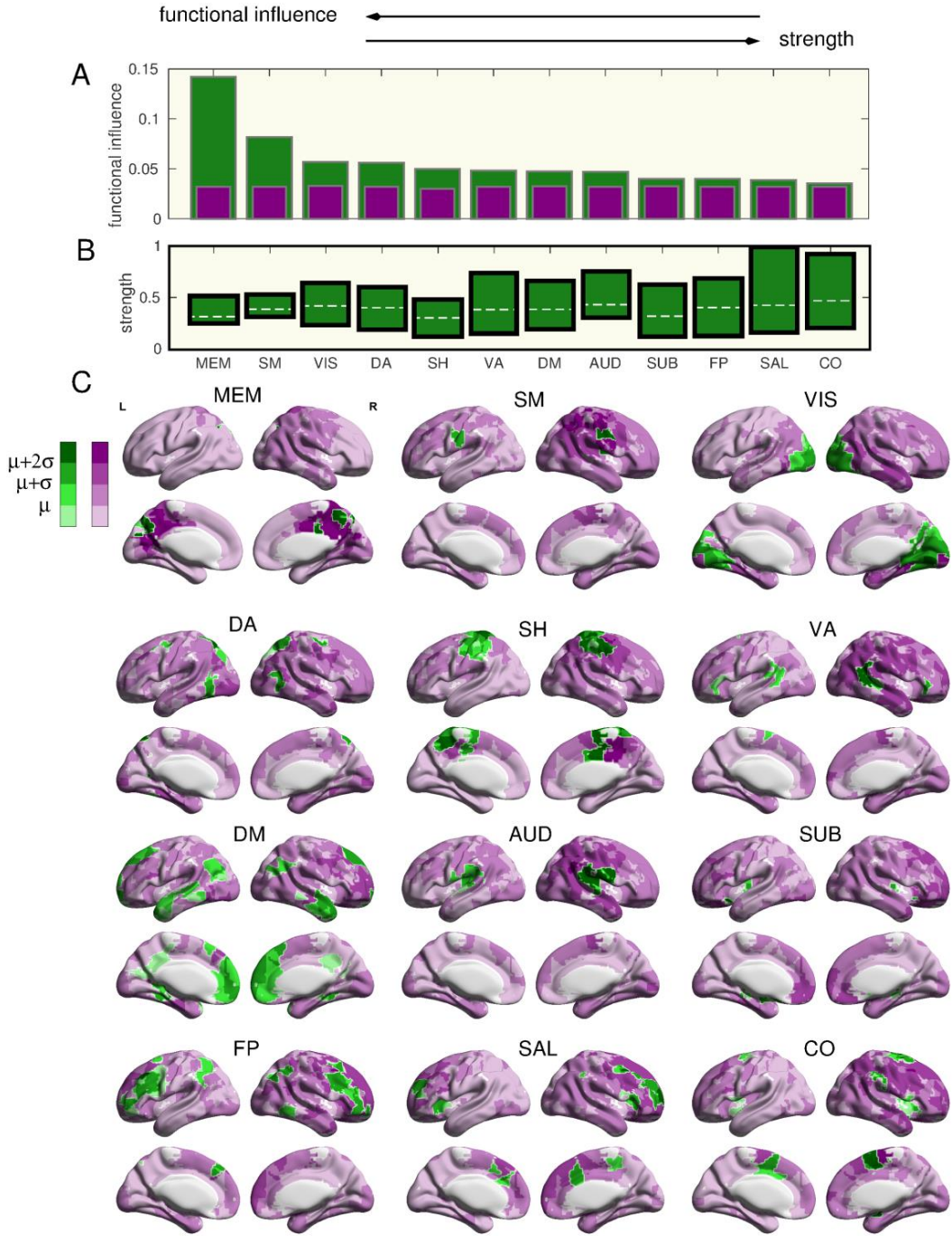

**Figure 7.** *Functional influence of brain systems*. A. Functional systems ranked by their functional influence, see text for details ($\Delta\omega = -0.002$ Hz). Green bars refer to the average changes in functional connectivity *within* a system following stimulation of each region *within* the system (i.e., impact of internal stimulation on system dynamics). Purple bars refer to the average change in functional connectivity *within* the system of interest following stimulation of all regions *outside* the target system (i.e., the impact of external stimulation on a given functional system). B. Anatomical connectivity strength of each functional system. Green bars indicate the minimum and maximum strength of regions defining the system. The dashed lines represent the mean strength. C. Spatial map of changes in functional connectivity within each system. Darker colors represent regions that, when stimulated, cause larger changes in functional connectivity within the system of interest. Green indicates the functional influence of stimulated regions within a given system whereas purple

indicates the functional influence of regions outside a given system. MEM = Memory system; SM = Somatosensory Mouth; VIS = Visual; DA = Dorsal Attention; SH = Somatosensory Hand; VA = Ventral Attention; DM = Default Mode Network; AUD = Auditory; SUB = Subcortical; FP = Fronto Parietal; SAL = Salience and CO = Cingulo Opercular.

To further understand the effects of local stimulation on the whole brain, we also analyzed the impact of local regional inhibition or excitation on functional connectivity in the systems of interest. In this last analysis we assess the impact that local inhibition of neural activity may have on the system connectivity as a function of the position of the stimulated region in the cortical hierarchy. Here we define the region's position within the hierarchy using the anatomical connectivity strength. We stimulate two representative regions within each system taken from opposite ends of the system-specific hierarchy (i.e., the top hub region and bottom peripheral region). Figure 8 presents the effects of local changes in brain activity in the top hub and bottom peripheral region of the visual (Fig. 8A) and cingulo-opercular systems (Fig. 8B). Overall, we find that inhibition of hub regions results in weaker changes in $\Delta|FC|$ within the system of interest compared to the same inhibition of peripheral regions (see brains in Fig. 8). These results are consistent with the tuning curves presented in Fig. 6. In fact, both the variability across regions as well as the magnitude of the effect of stimulation is enhanced in peripheral regions.

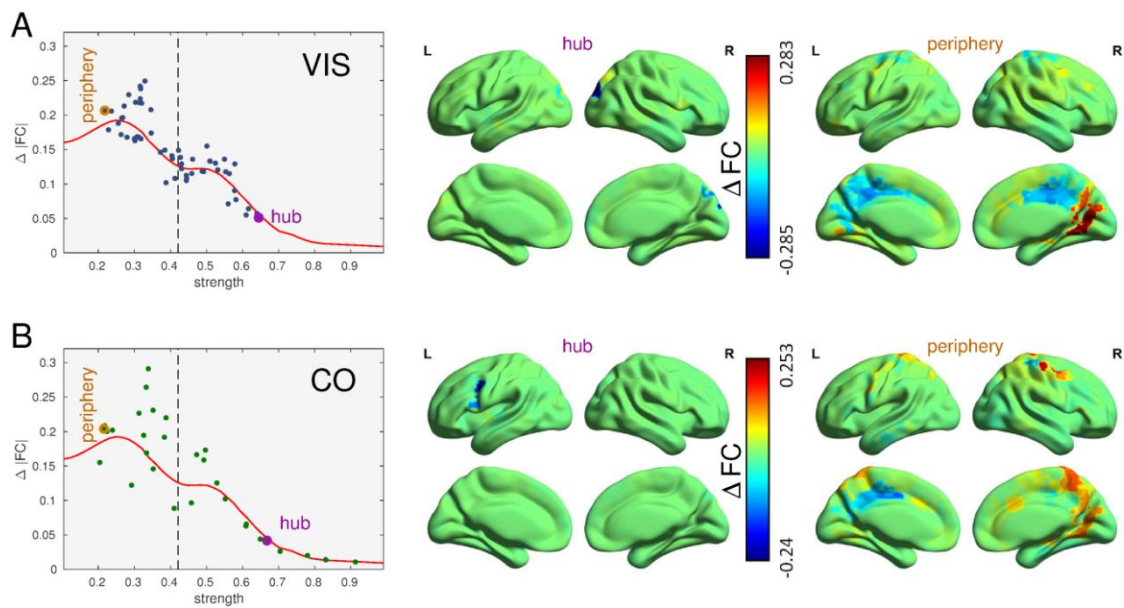

**Figure 8**. *Comparison between the effects of stimulating hub and periphery regions within a defined large-scale system.* A. Tuning curve (left) and system-level effects caused by an inhibitory stimulation ($\Delta\omega = -0.0008$ Hz) of a hub and a peripheral region within the Visual system. The red line in the left

plots corresponds to inhibitory tuning smooth curve of Fig. 6A. B. Same as panel A but for the Cingulo Opercular system.

**Discussion**

In the first part of this paper we briefly reviewed empirical and computational studies of how local changes in brain activity may affect patterns of functional connectivity across the whole brain. The overview of the existing literature highlights the potential of combining neuroimaging, local non-invasive brain stimulation and computational modeling to map and understand the neural principles supporting the propagation of local changes in brain activity across systems of the brain. In the second part of this paper we reported an original study, where we combined fMRI, MR diffusion tractography, and computational modeling to comprehensively map the effects of local inhibition or excitation on the whole brain network and large-scale brain systems. Results reveal key anatomical and functional principles determining the causal impact that a local perturbation in neural activity has on large-scale patterns of functional connectivity. We found that the system effects of local perturbations followed a relatively smooth tuning curve as a function of the regional anatomical connectivity strength. Specifically, our results indicate the importance of a core-periphery axis in shaping brain dynamics. This study also represents a first step toward a functional atlas mapping the impact of local stimulation on large-scale brain systems. This atlas will have important practical applications such as guiding the targeted use of local TMS to restore large-scale brain systems activity (Sale et al., 2015).

*Stimulation tuning curves emerge from a structural and functional hierarchy*

Our results suggest that the strength of the anatomical connectivity, and the related frequency of oscillations, is a key element for predicting the influence that perturbations to distinct brain regions have on large-scale neural systems. By plotting the influence of local stimulation on functional connectivity against the anatomical connectivity strength, we revealed tuning curves governing the brain's TMS-induced responses. The behavior of the tuning curves appears to generalize across the different functional systems (Fig. 7), suggesting that the tuning curves represent a fundamental property of brain dynamics. Notably, stimulation of peripheral regions, which are only weakly connected with other brain regions, causes larger changes in functional connectivity both at the whole-brain (asymmetric shapes of tuning curves in Fig. 6) and the system-specific levels (Fig. 7) compared to stimulation of strongly connected core regions (hubs). These results reflect the fact that peripheral regions exhibit more flexible dynamics whereas the dynamics of hub regions is more stable and constrained (Gollo et al., 2015).

The greater impact of peripheral regions on whole brain functional connectivity can be explained by the fact that such regions are dominant in the brain (Fig. 1B). Thus, although weakly connected, these regions have small frequency differences between themselves and a large number of other regions. Therefore, a small change in the activity of a peripheral region affects a large set of regions that operates at similar timescales. Conversely, a region that exhibits a very different intrinsic activity compared to other regions of the brain is more likely to be indifferent to the small variations in dynamics following local stimulation. These findings are consistent with empirical observations showing that inhibition of peripheral regions such as the right motor cortex engenders a significant pattern of changes in connectivity between this region and regions within and outside the sensorimotor systems (Cocchi et al., 2015). On the other hand, stimulation of a core region such as the FEF causes selective changes within the visual system (Cocchi et al., 2016; Ruff et al., 2006). Our findings also suggest that to achieve comparable systems-level effects in terms of magnitude, the intensity of the local stimulation needs to be tuned to the activity profile of the targeted region. In general, we found that the stimulation most likely to generate a significant change in large-scale functional connectivity is a strong inhibition of a peripheral region. This prediction appears in line with what is observed experimentally (Cocchi et al., 2015). Results from our work may therefore assist in the interpretation of behavioral effects of multi-regional TMS in the absence of imaging data.

The essential dynamics of most regions can be summarized by two parameters: the mean natural frequency ($\langle\omega_0\rangle$) and each region's detuning away from this mean. For example, in the case of inhibitory stimulation, peripheral nodes working at a higher frequency relative to the mean tend to increase their functional connectivity with the rest of the brain. In contrast, core (hub) regions with slower dynamics decrease their functional connectivity after stimulation (Fig. 3A). As illustrated in Fig. 4, this particular example has recently been confirmed by experiments in which one peripheral region (V1/V2) and a core region (FEF) were selectively inhibited in healthy human participants (Cocchi et al., 2016). Therefore, the tuning curves presented in Fig. 3 provide a general framework to predict the system-level effect of focal brain stimulation.

When we focused on the changes in functional connectivity induced by local inhibition and excitation we found that these effects are symmetric, yet opposite (Fig. 3). Local inhibition reduces the natural frequency of faster peripheral regions, bringing them closer to the other brain regions and typically increases the functional connectivity between the stimulated region and the rest of the brain. In contrast, local excitation increases the natural frequencies of these already-faster regions, which saturates for larger intensity, and hence tends to have a weaker effect on global functional connectivity. The converse occurs for core regions because these regions are slower than the typical oscillatory frequency. Further reducing a core region's oscillatory frequency (inhibition) usually disconnects the region from other regions of the brain, while an increase in the frequency (excitation) increases the stimulated region's functional connectivity with other regions. Excitation and inhibition also have similar effects on brain dynamics in term of absolute changes in functional connectivity (Fig. 6A). Peripheral regions cause large effects because their natural frequencies are similar to the frequencies of a larger number of regions (Fig. 1B).

*Selectively manipulating large-scale functional brain systems via local stimulation*

Although functional connectivity is a fairly simple time-averaged measure of brain dynamics, its application to functional magnetic resonance imaging data has allowed the identification of distinct large-scale brain systems supporting the bulk of brain functions including external attention and introspection (Fox et al., 2005; Greicius et al., 2003; Power et al., 2011). In the past few years several groups, including our own (Cocchi et al., 2016; Cocchi et al., 2015), have assessed the impact that a focal perturbation in neural activity could have on the patterns of functional connectivity in specialized large-scale brain systems (Eldaief et al., 2011; Fox et al., 2014; Fox et al., 2012b; Grefkes and Fink, 2011; Sale et al., 2015; Watanabe et al., 2013). Here we have extended these empirical works by quantifying how local stimulation across the whole brain affects the intrinsic dynamics of these brain systems.

Our results showed that the functional influence, defined as the mean effect of local stimulation on functional connectivity within a specific functional system, is constant for regions outside the targeted network. This effect appears independent of the system targeted by stimulation (purple in Fig. 7A). Conversely, the functional influence varies substantially across functional systems when the stimulation site lies within the targeted system (green in Figure 7A). Overall, these results are important because they support the notion that selective changes in the activity of spatially defined large-scale brain systems can be achieved using targeted, and appropriately calibrated, modulations in local neural activity (Sale et al., 2015).

Our findings reveal the existence of cortical hotspots that, if stimulated, have the potential to significantly alter the connectivity patterns of functional systems. Such hotspots are typically relatively large and do not have sharp boundaries (Fig. 7C). In fact, the functional influence decays smoothly (and not discontinuously) as we move to regions near the hotspots. This finding has some practical implication for TMS experiments, suggesting that small imprecisions in the coil position should not cause excessive differences in the effect of stimulation. A common target for TMS, identified here as a hotspot, is the dorso-lateral prefrontal cortex (DLPFC). This region is often targeted because it supports a broad range of key cognitive functions (Cieslik et al., 2012; Cocchi et al., 2013a; Cocchi et al., 2013b; Cole et al., 2010; MacDonald et al., 2000), its dysfunction is often related to psychiatric disorders (Cocchi et al., 2014; Fox et al., 2012a; Rajkowska et al., 2001; Weinberger et al., 1986), and it is easily accessible by TMS. In agreement with findings of Fox and colleagues (Fox et al., 2012a), our results suggest that the left DLPFC is a valuable stimulation target to modulate connectivity with and within the default mode brain network (Fig. 7C, DMN). Likewise, our simulations indicate that a greater effect on connectivity patterns between the left DLPFC and default mode regions is obtained by targeting the rostro-dorsal portion of the DLPFC (Fig. 7C). Our results suggest, however, that the right DLPFC provides an overall stronger modulation of functional connectivity with and within the default-mode brain system. This agreement between independent empirical and modeling results provides further support for the validity of our model. The maps of the functional influence of each region (Fig. 7C) may therefore assist the planning of future studies aiming at modulating specific functional systems using local TMS.

Our findings also indicate that functional systems can be classified based on their internal functional influence. In general, we found that peripheral systems such as the visual network and the somatosensory networks are more reactive to local changes in regional activity compared to high-level associative systems such as the cingulo-opercular network (Fig. 7A). Hence, local stimulation is more likely to change the patterns of connectivity of a system when the targeted brain region lies within a peripheral system than a high-level system. This result implies that the intensity of the stimulation needs to be adjusted according to both the targeted region and the targeted system. Such a prediction will need to be tested by future empirical work assessing the impact of local TMS as a function of the targeted system.

*From data to theory, and back again*

Our study highlights the potential of combining different empirical and theoretical tools to advance mechanistic knowledge on brain functions. In general, experiments are the guides and motivations for developing models and theories (Fig. 9). Valuable models and theories generate predictions for follow-up experiments aiming at validating and refining the explanation of the underlying mechanisms. Predictions (highlighted by the red box in Fig. 9), if correct, will bolster the strength of the model whereas incorrect predictions will guide the development of the next generation of models. More specifically, the results and tuning curves presented herein are predictions that will need to be tested in future experiments aiming to test and refine the proposed mechanisms underpinning system-level changes following focal perturbations. This link between model-driven predictions and experiments (red arrow in Fig. 9) allows closing the loop between experiment and theory.

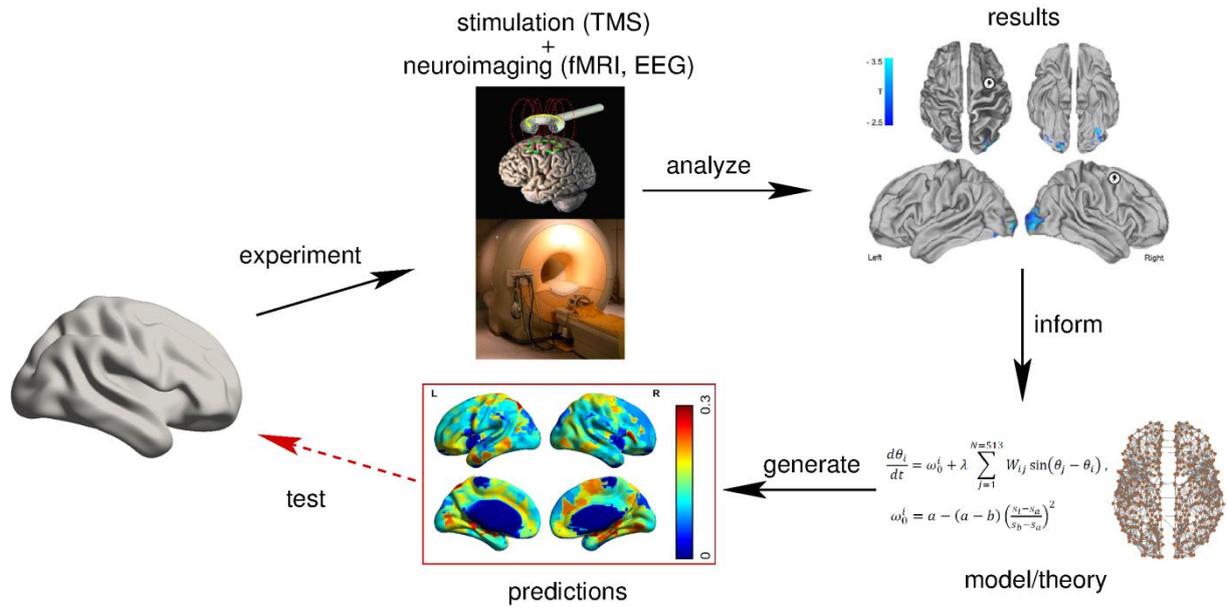

**Figure 9.** *Schematic representation of the iterative approach fusing theory and experiments.* Experiments are performed using a combination of stimulation and neuroimaging methods. These results, example extracted from Cocchi et al. (2016), inform models and theories that generate specific predictions on the underlying mechanisms supporting the observed effects. The testing of these predictions (red arrow) closes the loop and guides the next set of hypothesis-driven experiments. These results either provide evidence for the model or provide the impetus for an improved theory.

Although our study targeted a comprehensive set of regions, it relies on factors that may limit the generalization of our findings. The model uses a structural connectivity matrix that was derived from 75 healthy young adults (Roberts et al., 2016). Hence, the quality of our connectivity matrix depends on the accuracy of the underlying neuroimaging and tractography methods. Moreover, this connectome reflects an average human brain, and thus misses the distinctive aspects of each subject. It is possible that the key to understanding the large variability observed in the effects of brain stimulation might reside in these individual characteristics. In addition, the proposed model can only be considered experimentally validated for two (out of 513) regions: V1/V2 and FEF (Cocchi et al., 2016) The effects of stimulation over all other regions are predictions that are yet to be validated. A comprehensive mapping of the brain's stimulation-induced dynamics constitutes a major experimental endeavor for the near future.

*Conclusion*

In sum, our study highlights the potential to combine empirical and modeling tools to understand the neural principles supporting the system-level effect of focal cortical stimulation. Our findings emphasize the need to consider the intermingled functional and structural characteristics of targeted regions to predict the impact of local stimulation to the brain network. Specifically, our work demonstrates the importance of a periphery-core hierarchy to determine the effect of local stimulation on large-scale brain systems. By mapping such effects, this study provides novel resources to orient experiments aiming to test the mechanistic hypothesis generated by our model. These maps may also be valuable guides to find optimal stimulation sites to manipulate specific functional connections.

**Acknowledgments**

This work was supported by the Australian Research Council Centre of Excellence for Integrative Brain Function (ARC Centre Grant CE140100007) (JR) and the National Health and Medical Research Council (NHMRC): APP1110975 (LG), APP1099082 (LC).